\documentclass[final,3p,times]{elsarticle}

\usepackage{textcomp}

\usepackage{graphicx}
\usepackage{amssymb}
\usepackage{caption,amsmath,amsfonts}

\usepackage{lineno}

\date{31.01.2018}
\biboptions{compress}

\journal{Nuclear Instruments and Methods in Physics Research A}
\begin{document}

\begin{frontmatter}


\title{Quantitatively consistent computation of coherent and incoherent radiation in particle-in-cell codes - a general form factor formalism for macro-particles}



\author[HZDR,TUDD]	{R.~Pausch\corref{cor1}}

\ead{r.pausch@hzdr.de}
\cortext[cor1]{Corresponding author}
\address[HZDR]{Helmholtz-Zentrum Dresden - Rossendorf (HZDR), 01314 Dresden}
\address[TUDD]{Technische Universit\"at Dresden, 01062 Dresden}

\author[HZDR]			{A.~Debus}
\author[HZDR,TUDD]	{A.~Huebl}
\author[HZDR,TUDD]	{U.~Schramm}
\author[HZDR,TUDD]	{K.~Steiniger}
\author[HZDR]			{R.~Widera}
\author[HZDR]			{M.~Bussmann}

\begin{abstract}

Quantitative predictions from synthetic radiation diagnostics often have to consider all accelerated particles. 
For particle-in-cell (PIC) codes, this not only means including all macro-particles but also taking into account the discrete electron distribution associated with them.
This paper presents a general form factor formalism that allows to determine the radiation  from this discrete electron distribution
in order to compute the coherent and incoherent radiation self-consistently.
Furthermore, we discuss a memory-efficient implementation that allows PIC simulations with billions of macro-particles. 
The impact on the radiation spectra is demonstrated on a large scale LWFA simulation.

$\,$

\textit{ \textcopyright 2018. This manuscript version is made available under the CC-BY-NC-ND 4.0 license} 


\end{abstract}

\begin{keyword}
particle-in-cell simulations \sep laser plasma acceleration \sep far field radiation \sep plasma physics \sep radiation diagnostics


\end{keyword}

\end{frontmatter}

\hyphenation{PIConGPU}


\section{Particle-in-cell codes as synthetic diagnostic for laser plasma accelerators  }
\label{sec_intro}


Laser plasma-based accelerators such as laser wakefield accelerators (LWFA) offer various advantages over conventional accelerators such as their compact size due to their orders of magnitude larger acceleration gradient \cite{Tajima1979} resulting in $\mathrm{GeV}$ energy gains on centimeter scales \cite{Clayton2010,Wang2013,Leemans2014}, while allowing for compact electron bunch size \cite{Schnell2012,Kohler2016}, low beam emittance \cite{Geddes2012} and high current \cite{Couperus2017}. 
Since these laser plasma accelerators operate in a highly nonlinear regime, particle-in-cell (PIC) codes are essential for understanding the plasma dynamics in these experiments \cite{Gibbon2007,Esarey2009}.
Making quantitative predictions on the measurable outcome of these experiments is done by so-called synthetic diagnostics. 
An essential form of these synthetic diagnostics are predictions on the radiation emissions \cite{Hededal2005,Khachatryan2008,Sironi2009,Nishikawa2009,Martins2009,Nishikawa2011,Rykovanov2012,NISHIKAWA2012,Haugbolle2013,Nishikawa2013,Chen2013a,Pausch2014,NishikawaFermiSym2014,Pausch2014a,Martins2015,PauschPRE2017}.
However, most of these simulations lack quantitative predictions due to the small number of sample particles used (see \cite{Pausch2014} for details), but also due to an inconsistent treatment of the discrete nature of the electrons represented by macro-particles in PIC codes.
This often renders quantitative comparisons to experiments difficult to impossible. 

The general form factor formalism introduced in this paper overcomes this problem. 

\section{Spectrally resolved far field radiation using Li\'{e}nard Wiechert potentials}


The spectrally resolved far-field radiation emitted by a single electron can be computed by using Li\'{e}nard-Wiechert potentials \cite{Jackson1998}. 
Based on the electrons position $\vec{r}$, velocity $\vec{\beta}$ and acceleration $\dot{\vec{\beta}}$ over time, the energy emitted per unit frequency $\omega$ and unit solid angle $\Omega$ in direction of the unit vector $\vec{n}$ computes as:
\begin{equation} \label{theEquation}
\frac{\operatorname{d^2} I}{\operatorname{d} \Omega \operatorname{d} \omega}  =  \frac{q^2}{16 \pi^3 \varepsilon_0 c} \left| \int \limits_{-\infty}^{+\infty} \vec{A} \cdot \chi \operatorname{d} t   \right|^2 
\end{equation}
%
%
with 
$ \vec{A}  =  \frac{\vec n \times \left[ \left(  \vec n - \vec \beta  \right)  \times \dot{\vec \beta}  \right]}{\left( 1 - \vec \beta \cdot \vec n \right)^2}$ and 
$\chi   =  \text{e}^{\text{i} \omega (t - \vec n \cdot \vec r(t) / c)}$
being the radiation amplitude and complex phase while $\varepsilon_0$, $q$, and $c$  are the vacuum permittivity, the electron charge and mass.

For multiple electrons, the phase-relation between the various electrons needs to be taken into account by summation of the radiation amplitudes before taking the absolute square in Eq.~\ref{theEquation}:
\begin{equation} \label{theEquationMany}
\frac{\operatorname{d^2} I}{\operatorname{d} \Omega \operatorname{d} \omega}  (\omega, \vec{n}) = \frac{1}{16 \pi^3 \varepsilon_0 c} \left| \sum \limits_{k=1}^{N_e} q_k \int \limits_{-\infty}^{+\infty} \vec{A}_k \cdot \chi_k \operatorname{d} t   \right|^2 
\end{equation}
with $\vec{A}_k$ and $\chi_k$ being the radiation amplitude and phase of the $k$\textsuperscript{th} electron of all $N_e$ electrons considered. 
The charge $q_k$ of each electron has been moved inside the summation.
This combined radiation calculation leads to constructive and destructive interference depending on the electron distribution in phase space, the observed frequency, and the observation direction. 

\section{Applying form-factors to macro-particles}

\subsection{Macro-particle shapes in particle-in-cell codes}
Particle-in-cell codes discretize the particle distribution function of the Vlasov equation using a finite number of sample particles \cite{Hockney1988,Birdsall1991a}. 
Due to the approximately hundred billion electrons that need to be considered in a common LWFA simulation, the number of simulated particles is reduced by combining electrons into so-called macro-particles, that represent ensembles of electrons.
These macro-particles have a spatial charge distribution, but only a singular momentum to avoid a spatial separation of the charge distribution over time.
The commonly used current deposition algorithms in PIC codes treat the macro-particles as a continuous charge distribution \cite{Villasenor1992, Esirkepov2001}.
Using smoother and spatially more extended charge distributions for these macro-particles yields less numerical noise in the electromagnetic fields \cite{Esirkepov2001}. 
Such a higher order charge distribution, that works efficiently with the discretized fields in PIC codes, can be computed by convolving an existing charge distribution with a box function of spatial extent equivalent to a cell in the PIC code \cite{Hockney1988}
\begin{equation}
\rho^{(i+1)}(\vec{r}) = \rho^{(i)}(\vec{r}) \otimes \Pi(\vec{r}) \quad ,
\end{equation}
with $\rho^{(i)}(\vec{r})$ being the charge distribution of order $i$, and $ \Pi(\vec{r})$ being the box function.
Starting with a delta-distribution as a zeroth-order distribution $\rho^0  = \delta(\vec{r})$, higher order shapes can be derived (see sec.~\ref{sec_PICshapes}).  

\subsection{Form factors for arbitrary particle distributions}

In order to take into account the charge distribution of a single macro-particle, Eq.~\ref{theEquation} needs to include an integration over the charge distribution $\rho(\vec{r})$ analogous to the summation over all particles in Eq.~\ref{theEquationMany}.  

\begin{equation} \label{theEquationDistr}
\frac{\operatorname{d^2} I}{\operatorname{d} \Omega \operatorname{d} \omega} = \frac{1}{16 \pi^3 \varepsilon_0 c} \left| \int \limits_{V} \mathrm{d}V \rho(\vec{r}) \int \limits_{-\infty}^{+\infty} \vec{A} \cdot \chi \operatorname{d} t   \right|^2 
\end{equation}

\begin{figure}[!t]
	\includegraphics[width=0.45\textwidth]{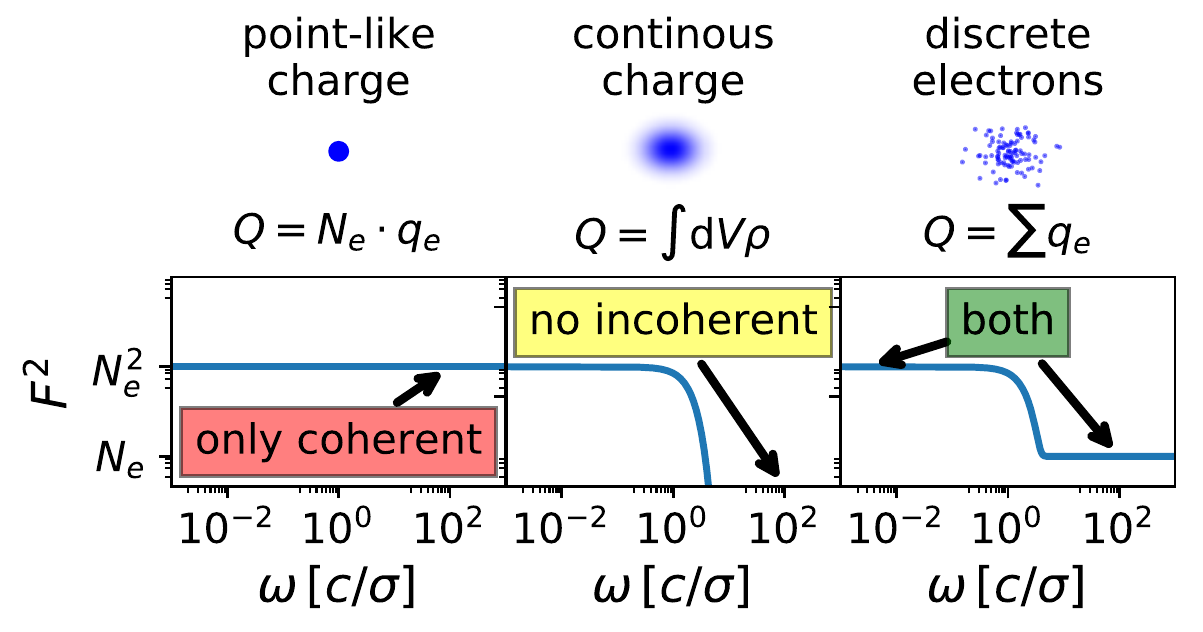}
    \caption[]{
    Various treatments of the macro-particle charge distribution and its influence on the incoherent radiation. 
    The discrete electron distribution results in both coherent and incoherent radiation. 
    }
	\label{picSchemeFF}
\end{figure}

Without loss of generality, we assume that the observation direction is parallel to the x-axis of the coordinate system $\vec{n} = \vec{e}_x$.
Furthermore, the charge distribution $\rho(\vec{r})$ can be assumed to be located around the position $\vec{r}_0 = (x_0, y_0, z_0)$. 
This allows integrating $\rho = \rho(x) \rho(y) \rho(z)$ over the other two axes spatially thus reducing the innermost spatial integration to:
\begin{eqnarray}
\int\limits_{-\infty}^{+\infty} \operatorname{d} x \rho(x) \chi & = & \int\limits_{-\infty}^{+\infty} \operatorname{d} x \rho(x) \text{e}^{\text{i} \omega (t - x / c)} \\
& = & \text{e}^{\text{i} \omega (t - \vec{n} \vec{r}_0 / c)}  \cdot \mathcal{F}[\rho(x-x_0) ] \quad ,
\end{eqnarray}
with $\mathcal{F}[\rho]$ being the Fourier transform of the charge distribution. 
For a point-like charge distribution, $\rho(\vec{r}) = \delta(\vec{r} - \vec{r}_0)$, one ends up with Eq.~\ref{theEquation}.
Since the entire charge is located in one point, the emitted radiation is completely coherent (Fig.~\ref{picSchemeFF}).

For higher order charge distributions, the additional factor 
from the Fourier transform leads to vanishing energies at high frequencies, so that any incoherent radiation at these frequencies is neglected (Fig.~\ref{picSchemeFF}).
As an example, the Fourier transform of the CIC particle shape with spatial extent $\Delta$ leads to $\mathcal{F}(\omega) = \Delta/c \cdot \mathrm{sinc}(\omega \Delta/2c )$, which results in vanishing intensities for frequencies $\omega \gg 2c/\Delta$, while for $0 \leq \omega < 2c/\Delta$, the radiation is similar to a point like charge distribution. 
Such particles would not radiate at frequencies above $\omega \gg c/\Delta$.
Obviously, this is physically not correct. 
In order to overcome this numerical artifact, the discrete nature of the electrons associated with a macro-particle needs to be taken into account. 

If the macro-particle represents $N$ electrons $\rho(\vec{r}) = \sum \limits_{k=1}^{N} q_e \delta(\vec{r} - \vec{r}_k)$, Eq.~\ref{theEquationMany} simplifies to
\begin{eqnarray} \label{Intro_Eq_theEquationMany_FF_01}
\frac{\operatorname{d^2} I}{\operatorname{d} \Omega \operatorname{d} \omega} 
& = &  \frac{q_e^2}{16 \pi^3 \varepsilon_0 c} \left| \int \limits_{-\infty}^{+\infty} \vec{A} \cdot \bar{\chi} \cdot \sum\limits_{k=1}^{N} \bar{\chi}_k\operatorname{d} t   \right|^2  \, \, ,
\end{eqnarray}
with
$\bar{\chi} = \text{e}^{\text{i} \omega (t - \vec n \cdot \vec {r}_0(t) / c)}$
being the phase of a central position $\vec{r}_0$ of the particle distribution and  
$\bar{\chi}_k = \text{e}^{\text{i} \omega \vec n \cdot \hat{\vec{r}}_k(t) / c} $
being the additional phase correction with the relative position $\hat{\vec r}_k(t) = \vec r_k(t) - \vec{r}_0(t)$. 
Since relativistic PIC codes assume the relative position to be stationary, one can separate the sum over all phase corrections and define a form factor $F^2$.
\begin{equation}\label{eqFF01}
F^2 (\omega) = \left|  \sum\limits_{k=1}^{N} \text{e}^{\text{i} \omega \vec n \cdot \hat{\vec{r}}_k  / c} \right|^2  \quad ,
\end{equation}
This allows to simplify Eq.~\ref{Intro_Eq_theEquationMany_FF_01} by separating the dynamic component of the macro-particle motion from its static charge distribution.
\begin{equation}
\frac{\operatorname{d^2} I}{\operatorname{d} \Omega \operatorname{d} \omega}  =  \frac{q_e^2}{16 \pi^3 \varepsilon_0 c} \left| \int \limits_{-\infty}^{+\infty} \vec{A} \cdot \bar{\chi} \operatorname{d} t   \right|^2 \cdot F^2(\omega)
\end{equation}
Since the exact distribution of electrons in a macro-particle is unknown, Eq.~\ref{eqFF01} cannot be evaluated exactly. 
The probability of a single electron with index $k$ being located at $\vec{r}_k$ is described by the probability density function $\psi(\vec{r}_k)$.
It is proportional to the continuous charge distribution of a macro-particle used in the PIC model $\psi(\vec{r}_k) \sim \rho(\vec{r}) $.
The probability density function of all $N$ electrons modeled by a macro-particle is the product of all individual distributions
$ \psi_N(\vec{r}_1, \vec{r}_2, \dots, \vec{r}_N) = \prod \limits_{k=1}^{N}  \psi(\vec{r}_k)$.
This combined probability density allows to define a form factor $F^2 (\omega)$ by averaging over all possible electron positions associated with a macro-particle.
Without loss of generality, $\vec{n} = \vec{e}_x$ can be assumed.
The average form factor can thus be computed by an $N$-dimensional integral. 
\begin{equation}
F^2 (\omega) \coloneqq \frac{\int \limits_{-\infty}^{+\infty}  \mathrm{d}x_1 \dots \int \limits_{-\infty}^{+\infty}  \mathrm{d}x_N \,  \rho(x_1, \dots, x_N)    \left|  \sum\limits_{k=1}^{N} \mathrm{e}^{i \omega x_k}   \right|^2}{  \int \limits_{-\infty}^{+\infty} \mathrm{d}x_1 \dots \int \limits_{-\infty}^{+\infty}  \mathrm{d}x_N  \, \rho(x_1, \dots, x_N)  }
\end{equation}
A lengthy but trivial integration over all particle positions yields the solution
\begin{eqnarray}
F^2(\omega) & = & N + \left( N^2 -N \right) \left| \int \limits_{-\infty}^{+\infty} \mathrm{d}x  \rho(x) \mathrm{e}^{\mathrm{i} \omega x} \right|^2 \\
 & = & N + \left( N^2 -N \right) \cdot \left(\mathcal{F}(\rho(x)) \right)^2 \label{theFFsolution}
\end{eqnarray}
where the first summand in Eq.~\ref{theFFsolution}  represents the incoherent and the second summand the coherent radiation of the $N$ electrons.

This equation was derived in \cite{Schiff1946} in the context of accelerator beam diagnostics for determining the bunch duration by measuring the coherent radiation cutoff (e.g. experimentally used in \cite{Hirschmugl1991, Lai1994, Carr2002, Abo-Bakr2002}). 
To the best knowledge of the authors, this is the first application of these form factors on macro-particles in PIC simulations.  
In contrast to these previous applications, in PIC simulations both the coherent and the incoherent regime are equally important and of interest as discussed in detail in section \ref{sec_LWFA}. 

\subsection{Integrating form factors in PIC codes}\label{sec_PICshapes}

In PIC simulations, the number of electrons represented by a macro-particle (given by the so-called weighting) can vary for each macro-particle. 
In order to evaluate Eq.~\ref{theEquationMany} numerically, as e.g. described in \cite{Pausch2014,Pausch2014a}, the following equation needs to be calculated:
\begin{equation}\label{Intro_Eq_theEq_multiple02}
\frac{\operatorname{d^2} I}{\operatorname{d} \Omega \operatorname{d} \omega}  =  \frac{\Delta t \cdot q_e}{16 \pi^3 \varepsilon_0 c} \left| \sum \limits_{j=0}^{N_t} \sum \limits_{m=1}^{N_\mathrm{mp}} F_m(\omega)  \cdot \vec{A}_{j,m} \cdot \chi_{j,m}  \right|^2   ,
\end{equation}
with the additional factor $ F_m(\omega)$ being the square root of the form factor of the $m^\mathrm{th}$ of all $N_\mathrm{mp}$ macro-particles at frequency $\omega$.
The integration over time is a sum over all $N_t$ discrete simulation time steps $\Delta t$. The radiation amplitude $\vec{A}_{j,m}$ and the complex phase $\chi_{j,m}$ are defined at the discrete time $t = j \cdot \Delta t$.

The most memory-efficient approach to computing this equation within the particle-in-cell framework is to evaluate the inner sum for each time step.
This requires applying the form factor to each particle before summation.
If memory is not a limitation, the form factors can be applied after the time integration to each particle in a post-processing manner. 
This is however only technically feasible for a small number of test particles, not for billions of macro-particles commonly used in simulations. 
Applying the form factor in-situ during the simulation to the inner summation thus allows handling billions of macro-particles. 
\begin{table}[!t]
\centering
\caption{List of common macro-particle shapes $\rho(x)$ in PIC codes and their associated form factor $F^2(\omega)$.}
\label{tab_FF}
\begin{tabular}{|c|c|c|}
\hline
\textbf{name}  & $\mathbf{\rho(x)}$                                  & $\mathbf{F^2(\omega)}$                                                            \\ \hline 
CIC   & $\Pi(x)$                                   & $\left( \frac{\Delta}{c} \cdot \mathrm{sinc}\left( \frac{\omega \Delta}{  2c} \right) \right)^2$ \\ 
TSC   & $\Pi(x) \otimes \Pi(x) = \Lambda(x)$       & $\left( \frac{\Delta}{c} \cdot \mathrm{sinc}\left( \frac{\omega \Delta}{  2c} \right) \right)^4$ \\ 
QSC   & $\Lambda(x)  \otimes \Pi(x)$               & $\left( \frac{\Delta}{c} \cdot \mathrm{sinc}\left( \frac{\omega \Delta}{  2c} \right) \right)^6$ \\ 
Gauss & $\mathrm{exp}\left( - \frac{x^2}{2 \sigma^2} \right) $ & $\frac{\sigma^2}{c^2} \cdot \mathrm{exp}\left( - \frac{\omega^2 \sigma^2}{ c^2} \right)$                      \\ \hline
\end{tabular}
\end{table}

The common macro-particle shapes in PIC codes are cloud-in-cell (CIC), triangular-shaped density cloud (TSC) and quadratic-spline density shape (QSC) \cite{Hockney1988}.
Their shape and associated form factor $F^2$ are listed in table~\ref{tab_FF}, with $\Pi(x)$ being the rectangular function of width $\Delta$.
Due to the convolution theorem, the form factor of these shapes is a higher power of the form factor of the CIC shape.
They all contain side lobes due to the finite cut-off of the charge distribution. 
A distribution without side lobes is a Gaussian charge distribution, also listed in table~\ref{tab_FF}.
It 
can thus be considered more suited for exploring radiation signatures where particle-shape induced peaks should be avoided.
For implementation in radiation damping algorithms, the self-consistent treatment with equivalent shapes as used in the PIC algorithm should be used.

\section{Coherent and incoherent radiation from laser wakefield acceleration}\label{sec_LWFA}

\begin{figure}[!t]
	\includegraphics[width=0.45\textwidth]{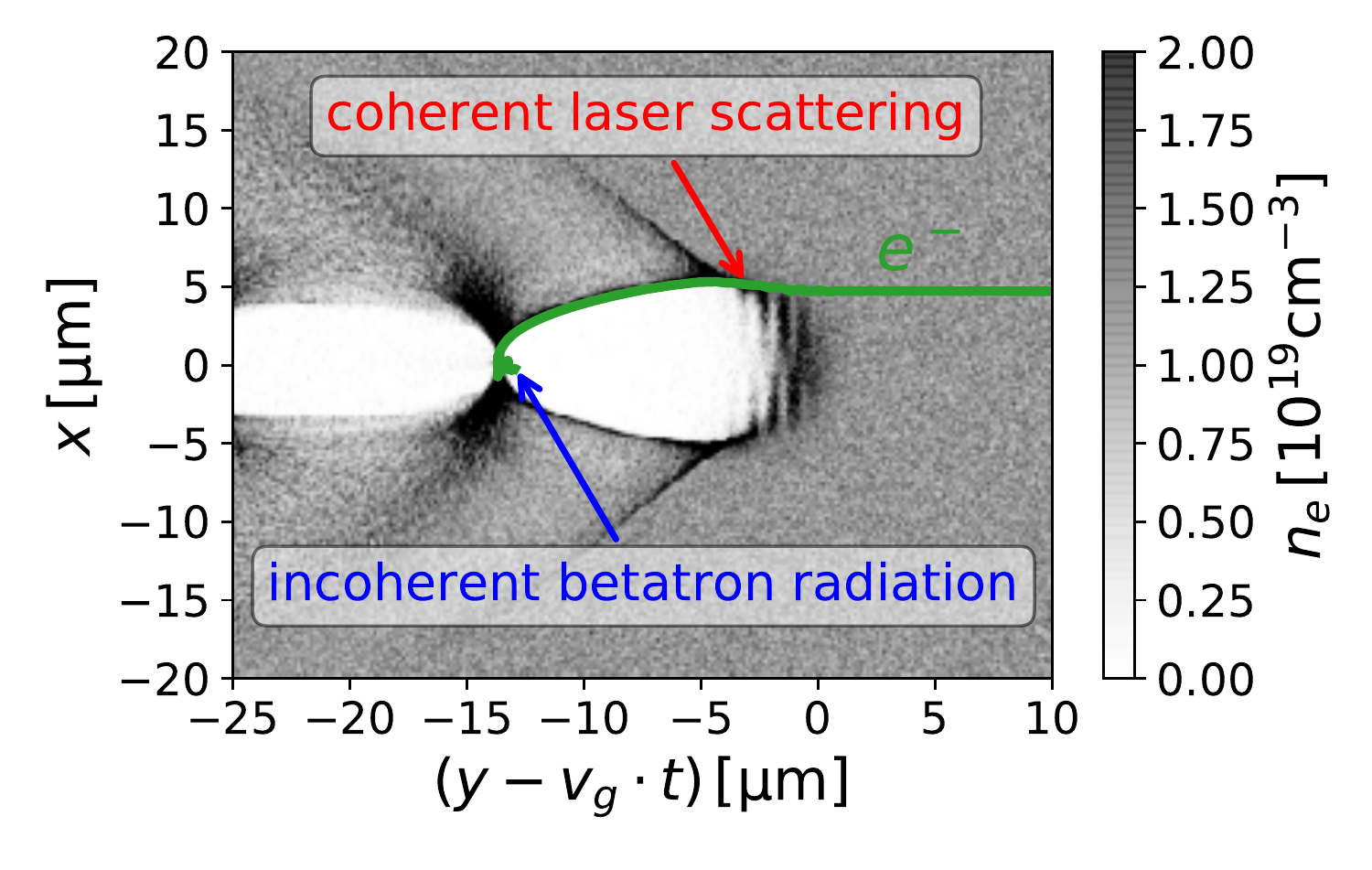}
    \caption[]{
    The density distribution during the self-injection is plotted alongside the entire trajectory of the macro  particle selected for radiation calculation. (LWFA simulation with laser: peak field strength $a_0=5$, duration $\tau=30\,\mathrm{fs}$, spot size $w_0 = 10 \, \mathrm{\mu m}$ in $n_e=10^{19}\, \mathrm{cm^{-3}}$ plasma density)
    The full configuration of the PIConGPU simulation presented can be found in \cite{ Pausch2018}.
    }
	\label{picLWFAdens}
\end{figure}
\begin{figure}[!t]
	\includegraphics[width=0.45\textwidth]{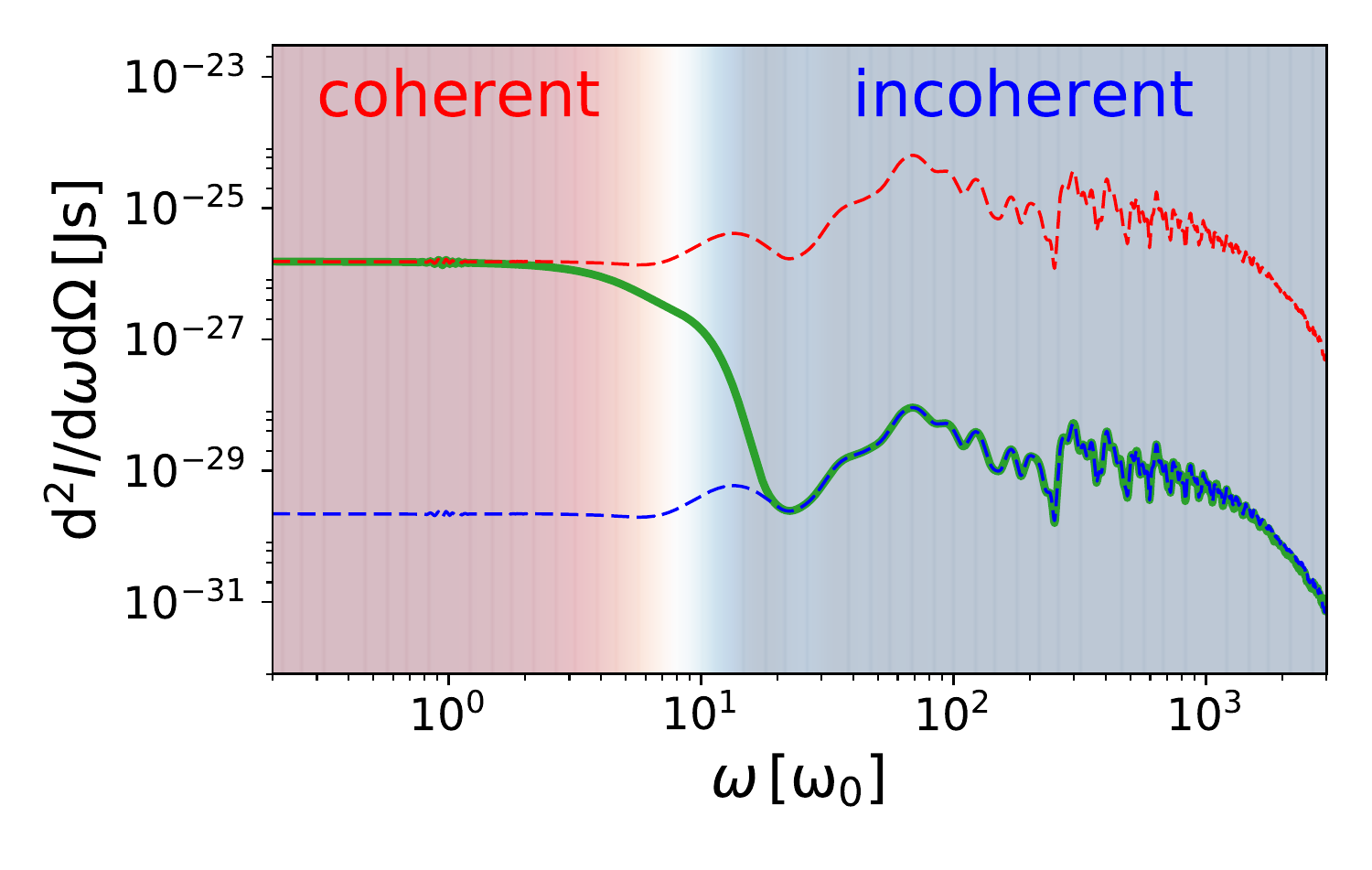}
    \caption[]{
The radiation spectra emitted by the sample macro-particle during an LWFA simulation. Using a form factor (bold line) allows to quantify the transition between coherent (upper dashed line) and incoherent scaling (lower dashed line). 
    }
	\label{picLWFAdrad}
\end{figure}

In this section, an LWFA simulation performed with the PIC code PIConGPU \cite{Burau2010, Bussmann2013} is presented to illustrate on a single macro-particle trajectory that both coherent and incoherent radiation is emitted by the same particle
and that neglecting the scaling of both regimes leads to overestimation of the incoherent radiation.
In this simulation, a plasma cavity is created into which macro-electrons are injected via self-injection. 
Fig.~\ref{picLWFAdens} illustrates the plasma dynamic during injection and the trajectory of the sample macro-particle for the entire simulation time in a co-moving frame.
%
During its entire dynamic, the macro-particle emits radiation at various frequencies (Fig.~\ref{picLWFAdrad}). 
The scattering of laser light is coherent while the betatron radiation in the x-ray frequency range \cite{Kohler2016} is incoherent. 
%

From the simple assumption that radiation at wavelengths shorter than the mean electron distance is incoherent, whereas radiation at wavelengths greater than the mean electron distance is coherent, it can be concluded that the transfer between coherent to incoherent radiation occurs at approximately $\omega \sim c/(2\sqrt[3]{n_e}) \approx 15 \cdot \omega_0$. 
This assumption neglects density accumulations shorter than this transition wavelength and applies only in a first approximation. 
The spectrum from the sample macro-particle is depicted in Fig.~\ref{picLWFAdrad} for a purely-coherent, purely-incoherent and a form-factor corrected scaling.
The simulated transition between coherent and incoherent radiation occurs at  $\omega \sim 10 \omega_0$ and thus agrees well with the simple estimation. 
Not correcting the spectra leads to an overestimation of the energy emitted at high frequencies.
As explained in section 3, the exact form of transition depends on the selected form factor. This should be kept in mind when comparing to experiments.

A variety of benchmarks tests of the radiation calculation in PIConGPU, based on relativistic and non-linear Thomson scattering, show excellent agreement with theoretical predictions \cite{ Pausch2014}.

We again stress that due to the $0.13$ billion macro-particles considered in this simulation, a post-processing application of the form factor would be technically intractable.


\section{Summary}

In this paper, we derived form factors for commonly used macro-particle shapes that allow to accurately quantify the amount of coherent and incoherent radiation in synthetic diagnostics efficiently and discussed an implementation in PIC codes. 
We demonstrated the need for this more thorough treatment of the discrete nature of the electrons represented by macro-particles in PIC codes on an LWFA simulation.

Using these form factors in synthetic radiation diagnostics not only paves the way towards quantitative radiation predictions from PIC codes that allow direct comparison with experiments, but also allows to predict the brightness of plasma based light sources such as betatron radiation or high harmonic generation. 


\section*{Acknowledgments}
This project is fully supported by the Helmholtz association under program Matter and Technology, topic Accelerator Research and Development. 
Computations were performed at the Center for Information Services and High Performance Computing (ZIH) at TU Dresden. 
A.H. has received funding from the European Unions Horizon 2020
research and innovation program under grant agreement No 654220.







\bibliographystyle{model1-num-names}
\bibliography{bibliography.bib}







\end{document}